\documentclass[floatfix,10pt,onecolumn,showpacs,amsmath,amssymb]{revtex4}

\usepackage{epsf}
\usepackage{graphicx}  
\usepackage{dcolumn}   
\usepackage{bm}        

\newcommand{\be}{\begin{equation}}
\newcommand{\en}{\end{equation}}
 \newcommand{\bea}{\begin{eqnarray}}
 \newcommand{\ena}{\end{eqnarray}}
  \newcommand{\sch}{Schwarzschild}

\begin{document}

\title{Misner-Sharp Mass and the Unified First Law in Massive Gravity}
\author{Ya-Peng Hu$^{1,3,4}$ \footnote{Electronic address: huyp@nuaa.edu.cn},Hongsheng Zhang$^{2,3~}$\footnote{Electronic address: hongsheng@shnu.edu.cn}}
\affiliation{ $^1$ College of Science, Nanjing University of Aeronautics and Astronautics, Nanjing 210016, China\\
$^2$ Center for Astrophysics, Shanghai Normal University, 100 Guilin Road, Shanghai 200234, China \\
$^3$ State Key Laboratory of Theoretical Physics, Institute of Theoretical Physics, Chinese Academy of Sciences, Beijing, 100190, China \\
$^4$INPAC, Department of Physics,and Shanghai Key Laboratory of Particle Physics and Cosmology, Shanghai Jiao Tong University, Shanghai 200240, China
 }


\begin{abstract}
We obtain the Misner-Sharp mass in massive gravity for a four dimensional spacetime with a two dimensional maximally symmetric subspace via the inverse unified first law method. Significantly, the stress energy is conserved in this case with a  general reference metric. Based on this property we confirm the derived Misner-Sharp mass by the conserved charge method.  We find that the existence of the Misner-sharp mass in this case does not lead to extra constraint for the massive gravity, which is notable in modified gravities. In addition, as a special case, we also investigate the Misner-Sharp mass in the static spacetime. We make a concise discussion of the stability problem in frame of the massive gravity in consideration. Especially, we take the FRW universe into account for investigating the thermodynamics of the massive gravity. The result shows that the massive gravity can be in thermodynamic equilibrium, which fills in the gap in the previous studies of thermodynamics in the massive gravity.

\end{abstract}


 \maketitle

\section{Introduction}

The mass (energy) of gravity field is a notorious problem. On the one hand, gravity field must be associated with mass, else gravitational wave becomes meaningless.  And the gravitational collapse will be inscrutable, since the energy of the matter of the original star vanishes.  A reasonable theory must ensure that gravitational wave carries mass and the spacetime relic after collapsing hides  energy equaling to that of the original star. On the other hand, a diffeomorphism invariant stress energy is strictly forbidden by the equivalence principle. Due to this principle, any free falling observer should sense a Minkowski spacetime, whose gravitational stress energy should be zero, and hence the gravitational stress energy should be zero for any observer according to the diffeomorphism invariance. This embarrassed status of stress energy of gravity forces us to consider quasi-local mass of gravity. We now have several different forms of quasi-local masses, for a review see \cite{qloc}.

In the spherically symmetric spacetime, one of the leading form is the Misner-Sharp mass. Misner and Sharp find some clues of the form of the Misner-Sharp mass by exploring the mass transformation from matter to gravitational field in a collapsing model  \cite{ms}. The explicit form of Misner-Sharp mass is presented in \cite{CMV}. An isolated defined quasi-local quantity seems not very interesting. If one want to deal with problems such as the mass transfers from one region to the other region, for example in the case of the gravitational wave radiation; or from one form to the other form, for example in the problem of gravitational collapsing, one had better introduce a conserved current to understand these astrophysical processes. In fact,   the Misner-sharp mass is a component of a conserved current corresponding to the timelike Killing vector in static spherically symmetric spacetime and to the Kodama vector in a general spherically symmetric spacetime. Some generalizations of the original Misner-Sharp mass has been introduced in \cite{mae} for the Gauss-Bonnet gravity, in \cite{cai} for four dimensional $f(R)$ gravity, and in \cite{self1} for higher dimensional $f(R)$ gravity.

All these generalizations are based on a unified definition of the Misner-Sharp mass. In an $n$-dimensional (n$\geq$3) spacetime which permits three-type  $(n-2)$-dimensional maximally symmetric subspace, we define the Misner-Sharp mass $M_{ms}$ as follows,
\be
M_{ms}=-\int *(T_{ab}K^{b}), \label{con2}
\en
where $T$ denotes the stress energy of the matter fields,  $K$ labels the Kodama vector (which reduces to a Killing one in stationary spacetime) \cite{koda}, and a $*$ represents the Hodge duality. We see that in this form the gravitational mass is always smuggled by the matter masses. In a vacuum region, for example a shell dwelling at $r_1$ to $r_2$ ($0<r_1<r_2$) in the  Schwarzschild spacetime, the Misner-Sharp mass is zero. The Misner-Sharp mass does not change when more vacuum regions (without singularities) are included. This is a critical property of the Misner-Sharp mass. In fact, the Misner-Sharp mass is the total mass of matter and gravity, for a clarifying of this point, see \cite{self2}. By properly using this point, we can obtain some exact solutions with special symmetries via thermodynamic considerations \cite{self3}.

The massive gravity is an important generalization of Einstein gravity. With the requirement of diffeomorphism invariance the Einstein gravity is the unique one which corresponds to a massless spin-2 boson \cite{tony}, though this is not the history style by which the Einstein theory is derived. In principle, we are always be curious to ask that what will happen if we introduce a mass for this spin-2 boson. But, in opposite to our intuition based on the Proca equation, this is a very intricate problem \cite{massive}. Several investigations have been made in massive gravity. In \cite{deRham:2010ik,deRham:2010kj,Hinterbichler:2011tt}, a classical of nonlinear massive gravity theories has been proposed, in which the ghost field is absent~\cite{hassan,Cai:2012db}.  In this class of massive gravity, generally the stress energy is no longer conserved.  Recently, Vegh~\cite{Vegh:2013sk} found a nontrivial black hole solution with a Ricci flat horizon in four dimensional massive gravity with a negative cosmological constant~\cite{Hassan:2011vm}. He then find that the mass of the gravitons can play the same role as the lattices do for the holographic conductor counterpart:  the conductivity generally exhibits a Drude peak which approaches to a delta function in the massless gravity limit. The black hole thermodynamics of this class of massive gravity is studied in \cite{Cai2015}. Some holographic consequence of the effect of graviton mass in massive gravity has been investigated in~\cite{Blake:2013bqa,Davison:2013jba,Davison:2013txa,Adams:2014vza,Amoretti:2014zha}.

 An interesting problem in massive gravity is that the conservation of the stress energy will break down if we still require the theory is diffeomorphism invariant. We can construct a massive gravity in which the stress energy is still conserved, the cost is that we lose the general diffeomorphism invariance, though we can ensure the Poincare invariance as an ordinary field theory in Minkowski space. It seems that we cannot have our cake and eat it too. This fascinating problem may root in the essence of the mass form of the massive gravity. The Misner-Sharp mass is a component of a conserved current associated to the matter stress energy by definition (\ref{con2}). Because the stress energy of the matters in massive gravity may be not a conserved quantity, i.e., generally
     \be
     \nabla_a T^{ab}\neq 0,
     \en
     the corresponding Misner-Sharp mass defined in (\ref{con2}) may be not a conserved charge.   We will try to explore the Misner-Sharp mass for a diffeomorphism invariant massive gravity in a four dimensional spacetime with a two dimensional maximally symmetric subspace, which may be helpful to understand this stress energy conservation problem. We shall show that in this special case (\ref{con2}) is still a conserved charge, and is exactly equal to the result from the inverse first law method. Furthermore, the studies to thermodynamics of massive may be also helpful to clear the conservation problem of the stress energy.

     The stability problem is critical for massive gravity. For a general non-degenerated reference metric, it is shown that the ghost excitation can be removed by a reparametrization of the ADM decomposition. In this article we will show that for a singular reference metric in consideration, the ghost excitation is also suppressed.

 Black hole thermodynamics (more generally gravi-thermodynamics) significantly boosts our understandings of gravity theory. It even be treated as a critical probe to the quantum gravity theory. In the solutions of the Einstein theory, the global quantities such as mass, entropy, temperature, angular momentum, and charge exactly satisfy the first law in equilibrium thermodynamics. However, since in the first law for black hole the global quantities are involved, it is difficult to be used in the realistic astrophysical processes. In view of this situation, the unified first law is developed. In the unified first law, only quasi-local quantities are involved.  Therefore, it is no essential difficulties to apply it in a finite region. Both in the first law and the unified first law the spacetime are treated as an equilibrium thermodynamic system. This point is correct if we restrained in the Einstein gravity.  But when we consider modified gravities, is this equilibrium condition automatically satisfied? It is not a trivial problem, i.e., some modified gravities can not reach an equilibrium state. When we consider a thermodynamical system, an entropy production term is inevitable. According to some investigations, $f(R)$ theory may be such an example~\cite{Jac,eling}. We will show that there is no entropy production term in massive gravity theory. It really can be in equilibrium state. This presents a decent foundations for the previous studies of thermodynamics in massive gravity \cite{Vegh:2013sk,Cai2015,lihui}.

This paper is organized as follows. First, we give a simple review for the Misner-Sharp mass in some previous results in section II. In section III, we investigate the Misner-Sharp mass in the massive gravity for a four dimensional general spacetime with a two dimensional maximally symmetric subspace, which is used by both the inverse unified first law method and conserved charge method. As a special case, we also investigate the Misner-Sharp mass in the static spacetime in the section IV. In section V, for the simplicity, we just take the FRW universe into account to investigate the thermodynamical state for the massive gravity, and we show that the massive gravity can be in thermodynamic equilibrium. Finally, we give a brief conclusion and discussion in section VI.

\section{Misner-Sharp mass: some previous results}
The Misner-Sharp mass is a significant quasilocal mass of gravitational field, i.e., defined on a boundary of a given region in space-time. It is first proposed in Einstein gravity in a spherically symmetric spacetime. For the four dimensional spherically symmetric spacetime with metric
   \be
   ds^2=h_{ab}dx^adx^b+r^2d\Omega_2^2,
   \label{2dim}
   \en
   where $\Omega_2$ denotes a unit two-sphere, $h_{ab}$ are general functions which are independent on the inner coordinates of the two-sphere, and $a,~b$ run from 0 to 1. The Misner-Sharp mass is defined as \cite{Hayward1}
   \be
   M_{ms}=\frac{r}{2G}\left(1-h^{ab}\partial_a r\partial_b r\right),
   \label{defi}
   \en
   where $G$ denotes the Newton constant.
 With this definition of the Misner-Sharp mass, the unified first law \cite{Hayward1} reads,
  \be
  dM_{ms} = A \Psi_a dx^a + W dV,
  \label{uni}
\en
where $A = 4\pi r^2$ is the area of the sphere with radius $r$ and
$V=4\pi r^3/3$ is its volume, $W$ is called  work density defined
as $W= -h^{ab}T_{ab}/2$ and $\Psi$ energy supply vector,
$\Psi_a=T_a^{\ b} \partial _b r +W \partial r_a$, with $T_{ab}$
being the projection of the four-dimensional stress energy
$T_{\mu\nu}$ of matter in the normal direction of the
2-dimensional sphere.

The unified first law has been used in several different cases of Einstein gravity and in modified gravity \cite{Cai-Cao}.
  The logic in the unified first law is as follows: we first define the Misner-Sharp mass as shown in (\ref{defi}), and then we find that it satisfies the unified first law (\ref{uni}). However, in the modified gravity theories, we have no prior definition of the Misner-Sharp mass. Exactly the unified first law requires a mass in the LHS of (\ref{uni}). Thus we can define a mass in (\ref{uni}) as an extension of the Misner-Sharp mass in modified gravity theories by using the RHS of (\ref{uni}) \cite{cai}. We call it inverse first law method, or the integration method in previous literatures. In addition, in a general spherically symmetric spacetime, one can define a Kodama vector. The stress energy together with the Kodama vector usually leads to a conserved current, whose corresponding conserved
charge is just the Misner-Sharp mass in Einstein gravity. This is the conserved charge method, and has been proved to be equivalent to the inverse unified first law method in the Gauss-Bonnet gravity and $f(R)$ gravity.

\section{Misner-Sharp mass in the massive gravity}
In this section, we first explore the Misner-Sharp mass in the massive gravity by using the inverse first law method. Then, we consider the conserved charge method. Usually, the action of the massive gravity in an $(n+2)$-dimensional spacetime reads~\cite{Vegh:2013sk,Cai2015}
\begin{equation}
\label{actionmassive}
S =\frac{1}{16\pi G}\int d^{n+2}x \sqrt{-g} \left[ R +\frac{n(n+1)}{l^2} +m^2 \sum^4_i c_i {\cal U}_i (g,f)\right],
\end{equation}
where   $f$ is a fixed symmetric tensor, which is usually called the reference metric,
$c_i$ are constants \footnote{For a self-consistent massive gravity theory, all those coefficients  might be required to be negative if $m^2>0$. However, in this paper we do not impose this limit, since in AdS space, the fluctuations of some fields with negative mass square could  still be stable if the mass square
obeys corresponding Breitenlohner-Freedman bounds. },  and ${\cal U}_i$ are symmetric polynomials of the eigenvalues of the $(n+2)\times (n+2)$ matrix ${\cal K}^{\mu}_{\ \nu} \equiv \sqrt {g^{\mu\alpha}f_{\alpha\nu}}$:
\begin{eqnarray}
\label{eq2}
&& {\cal U}_1= [{\cal K}], \nonumber \\
&& {\cal U}_2=  [{\cal K}]^2 -[{\cal K}^2], \nonumber \\
&& {\cal U}_3= [{\cal K}]^3 - 3[{\cal K}][{\cal K}^2]+ 2[{\cal K}^3], \nonumber \\
&& {\cal U}_4= [{\cal K}]^4- 6[{\cal K}^2][{\cal K}]^2 + 8[{\cal K}^3][{\cal K}]+3[{\cal K}^2]^2 -6[{\cal K}^4].
\end{eqnarray}
The square root in ${\cal K}$ means $(\sqrt{A})^{\mu}_{\ \nu}(\sqrt{A})^{\nu}_{\ \lambda}=A^{\mu}_{\ \lambda}$ and $[{\cal K}]=K^{\mu}_{\ \mu}=\sqrt {g^{\mu\alpha}f_{\alpha\mu}}$ (to extract the roots of the components one by one and then to make summation).  The equations  of motion turns out to be
\begin{eqnarray}
R_{\mu\nu}-\frac{1}{2}Rg_{\mu\nu}-\frac{n(n+1)}{2l^2} g_{\mu\nu}+m^2 \chi_{\mu\nu}&=&8\pi G T_{\mu \nu },~~
\end{eqnarray}
where
\begin{eqnarray}
&& \chi_{\mu\nu}=-\frac{c_1}{2}({\cal U}_1g_{\mu\nu}-{\cal K}_{\mu\nu})-\frac{c_2}{2}({\cal U}_2g_{\mu\nu}-2{\cal U}_1{\cal K}_{\mu\nu}+2{\cal K}^2_{\mu\nu})
-\frac{c_3}{2}({\cal U}_3g_{\mu\nu}-3{\cal U}_2{\cal K}_{\mu\nu}\nonumber \\
&&~~~~~~~~~ +6{\cal U}_1{\cal K}^2_{\mu\nu}-6{\cal K}^3_{\mu\nu})
-\frac{c_4}{2}({\cal U}_4g_{\mu\nu}-4{\cal U}_3{\cal K}_{\mu\nu}+12{\cal U}_2{\cal K}^2_{\mu\nu}-24{\cal U}_1{\cal K}^3_{\mu\nu}+24{\cal K}^4_{\mu\nu}).
\end{eqnarray}

In this article, for simplicity and without loss of generality, we just consider the Misner-Sharp mass of a four-dimensional spacetime with a two dimensional maximally symmetric inner space in the above massive gravity, and the general metric ansatz can be
  \be
ds^{2}=-2e^{-\varphi(u,v)}dudv+r^{2}(u,v)\gamma _{ij}dz^{i}dz^{j}.
\label{metricn1}
\en
where $\gamma _{ij}$ is the metric on a two-dimensional
constant curvature space ${\cal N}$ with its sectional curvature
$k=\pm 1,0$, and the two-dimensional spacetime spanned by two null
coordinates $(u,v)$ and its metric are denoted as $({\cal T}, h_{ab})$. In addition, we take the following reference metric as in \cite{Cai2015},
\begin{equation}
\label{reference}
f_{\mu\nu} = {\rm diag}(0,0, c_0^2 \gamma_{ij} ),
\end{equation}
with $c_0$ being a positive constant. Thus the symmetric polynomials become
\begin{equation}
{\cal U}_1= \frac{2c_0}{r}, {\cal U}_2= \frac{2c_0^2}{r^2}, {\cal U}_3= 0, {\cal U}_4= 0,
\end{equation}
and hence the field equation takes its form as
\begin{equation}
R_{\mu\nu}-\frac{1}{2}Rg_{\mu\nu}+\Lambda g_{\mu\nu}+m^2 \chi_{\mu\nu}=8\pi G T_{\mu\nu}, \label{Eqn1}
\end{equation}
where $\Lambda=-\frac{3}{l^2}$, and
\begin{eqnarray}
&& \chi_{\mu\nu}=-\frac{c_1}{2}({\cal U}_1g_{\mu\nu}-{\cal K}_{\mu\nu})-\frac{c_2}{2}({\cal U}_2g_{\mu\nu}-2{\cal U}_1{\cal K}_{\mu\nu}+2{\cal K}^2_{\mu\nu}).
\end{eqnarray}
For the metric (\ref{metricn1}), the useful components in the field equation (\ref{Eqn1}) can be explicitly expressed as
\begin{eqnarray}
&& 8\pi G T_{uu} =  -2\frac{r_{,u}\varphi_{,u} + r_{,uu}}{r}, \nonumber\\
&& 8\pi G T_{vv} =  -2\frac{r_{,v}\varphi_{,v} + r_{,vv}}{r}, \nonumber\\
&& 8\pi G T_{uv} = \frac{(k+c_0^2c_2m^2) - \Lambda r^2 + 2r_{,v}r_{,u}e^\varphi+r(c_0c_1m^2+2r_{,uv}e^\varphi)}{r^2e^\varphi}, \label{EqST}
\end{eqnarray}

\subsection{Inverse unified first law method}
According to the inverse unified first method in Ref \cite{Cai2015}, similar to
the case of Einstein gravity (\ref{uni}), one assumes the
equations (\ref{EqST}) of gravitational field can be cast into the form
\begin{equation}
dM_{eff}=A\Psi_a dx^a +WdV, \label{2eq5}
\end{equation}
where $A=V_{k}r^2$ and $V =V_{k}r^3/3$ are
area and volume of the $2$-dimensional space with radius $r$, energy supply vector $\Psi $ and energy density $W$ are
defined on $({\cal T},h_{ab})$ as in the case of Einstein gravity, and $M_{eff}$ is signed as the generalized Misner-Sharp mass in the modified gravity if it exists. Hence the right hand side in (\ref{2eq5}) can be explicitly expressed as
\begin{equation}
A\Psi_a dx^a +WdV=A(u,v)du+B(u,v)dv,
 \label{2eq6}
\end{equation}%
where
\be
A(u,v)=V_k r^2 e^{\varphi}(r,_{u}T_{uv}-r,_{v}T_{uu}),
\en
\be
B(u,v)=V_k r^2 e^{\varphi}(r,_{v}T_{uv}-r,_{u}T_{vv}).
\en\label{ABnew}
With the equations in (\ref{EqST}), we can express $A$ and $B$ in
terms of geometric quantities as
\begin{eqnarray}
A(u,v) &=&\frac{V_k}{8\pi G}\Big\{r,_{u}[(k+c_0^2c_2m^2) - \Lambda r^2 + 2r_{,v}r_{,u}e^\varphi+r(c_0c_1m^2+2r_{,uv}e^\varphi)]\notag \\
&&+2re^{\varphi}r,_{v}(\varphi,_{u}r,_{u}+r,_{uu})\Big\},
\notag \\
B(u,v) &=&\frac{V_k}{8\pi G}\Big\{r,_{v}[(k+c_0^2c_2m^2) - \Lambda r^2 + 2r_{,v}r_{,u}e^\varphi+r(c_0c_1m^2+2r_{,uv}e^\varphi)]\notag \\
&&+2re^{\varphi}r,_{u}(\varphi,_{v}r,_{v}+r,_{vv})\Big\}.
\label{2eq9}
\end{eqnarray}%

Now we try to derive the generalized Misner-Sharp mass by
integrating the equation (\ref{2eq5}).  Clearly, if it is
integrable, the following integrable condition has to be satisfied
\begin{equation}
\frac{\partial A(u,v)}{\partial v}=\frac{\partial B(u,v)}{\partial
u}.
 \label{2eq10}
\end{equation}%
It is easy to check that $A$ and $B$ given in (\ref{2eq9}) indeed
satisfy the integrable condition (\ref{2eq10}). Thus directly
integrating (\ref{2eq5}) gives the generalizing Misner-Sharp
energy
\begin{eqnarray}
E_{eff} &=&\int A(u,v)du + \int [B(u,v) - \frac{\partial}{\partial v}\int A(u,v)du]dv \nonumber\\
&=& \frac{V_kr}{8\pi G}[(k+2e^\varphi r_{,u}r_{,v}) - \frac{\Lambda r^2}{3} + \frac{(c_1c_0m^2r+2c_2c_0^2m^2)}{2}]. \label{MSMG}
\end{eqnarray}%
Note that, here the second term in the first line of (\ref{MSMG})
in fact vanishes and we have fixed an integration constant so that
$E_{eff}$ reduces to the Misner-Sharp energy in Einstein gravity
when $c_1=c_2=0$.
In addition, the generalized Misner-Sharp mass in massive gravity
can be rewritten in a covariant form
\begin{eqnarray}
\label{2eq12}
E_{eff}&=&\frac{V_kr}{8\pi G}[\left(k-h^{ab}\partial_a r\partial_b r\right)- \frac{\Lambda r^2}{3} + \frac{(c_1c_0m^2r+2c_2c_0^2m^2)}{2}].\label{MSCOMG}
\end{eqnarray}%
which is the generalized Misner-Sharp mass in the massive gravity with four dimensional spacetime with a two dimensional maximally symmetric subspace.

\subsection{Conserved charge method}
Note that, another important property of Misner-Sharp mass is that it can be associated with the conserved charge corresponding Kodama vector in a spherically symmetric spacetime. One can extend the Misner-Sharp mass to Gauss-Bonnet gravity by using this property \cite{mae}, which is equivalent to the inverse unified first method \cite{cai}. Here we check whether the two methods are equivalent in the four-dimensional massive gravity case. The special significance of this confirmation is associated to conservation problem of the stress energy of the massive gravity. It is not clear whether the mass derived by the inverse unified first law method is a component of a conserved current. Generally the stress energy is not conserved in massive gravity. Hence, the charge defined in (\ref{con2}) may be not a conserved charge. However, we will verify that the stress energy is really conserved in our case.

  In Ref. \cite{koda}, Kodama vector is defined in a spherically symmetric spacetime as,
\be
K^{\mu }=-\epsilon ^{\mu \nu }\nabla _{\nu}r,
\label{kodama}
\en
 on $({\cal M},~g)$, where $\epsilon _{\mu \nu }=\epsilon _{ab}(dx^{a})_{\mu }(dx^{b})_{\nu}$, and $
\epsilon _{ab}$ is the compatible volume element to the metric $h$ on $({\cal T},~h_{ab})$. The Greek indexes $\mu,~\nu$ run from $0$ to $1$ in (\ref{metricn1}). By direct calculation, (\ref{kodama}) can be rewritten as,
\be
K^{\mu}=e^{\varphi}\left[r,_{v}\Big(\frac{\partial }{\partial
u}\Big)^{\mu} -r,_{u}\Big(\frac{\partial }{\partial
v}\Big)^{\mu}\right].
\label{kodama2}
\en
 Just like the case of a Killing vector, the Kodama vector induces a conserved current in Einstein gravity. Mimicking the case of Einstein gravity, we define
 \be
 J^{\mu }=-T^{\mu }{~}_{\nu}K^{\nu}.
  \label{Jmu}
 \en
 It is interesting that the stress-energy of matter field is conserved for the massive gravity in our case, i.e. the ansatz (\ref{metricn1}) and reference metric (\ref{reference})
 \be
 \nabla _{\mu}T^{\mu\nu}=0.
 \en
Furthermore, we find that the current $J^{\mu}$ is also conserved in this case,
  \be
   \nabla _{\mu}J^{\mu}=0.
   \en
Therefore, the associated conserved charge can be defined
\begin{equation}
Q_{J}=\int {}^*J.
 \label{3eq15}
\end{equation}%
 By using the line
element in (\ref{metricn1}) and equations in (\ref{EqST}), we obtain
\begin{eqnarray}
Q_{J}=\frac{V_kr}{8\pi G}[(k+2e^\varphi r_{,u}r_{,v}) - \frac{\Lambda r^2}{3} + \frac{(c_1c_0m^2r+2c_2c_0^2m^2)}{2}].
\label{QJ}
\end{eqnarray}
which is obvious that the conserved charge $Q_J$ is just $M_{eff}$ in (\ref{MSMG}). We reach the same result by different routes. Therefore, (\ref{QJ}) and (\ref{MSMG}) should be a reasonable generalization of Misner-Sharp mass in massive gravity. At least, it inherits two significant properties of Misner-Sharp mass in Einstein gravity: it satisfies the unified first law and corresponds to a conserved charge associating to the Kodama vector. For $c_1=c_2=0$ our result reduces the previous result in Einstein gravity \cite{cai}.

\section {static case} \label {A}
The spacetime (${\cal M},~ g$) becomes a stationary one if there exists a time-like Killing vector. Moreover, it is a static one if it permits an $n-2$-dimensional maximally symmetric submanifold (${\cal N},~\gamma$), since the metric can be written in a time orthogonal coordinates system,
  \be
  ds^2=-\lambda(r)dt^2+h(r)dr^2+r^{2}\gamma _{ij}dz^{i}dz^{j}.
  \label{metrics}
  \en
In principle, for any concrete metric form one can always make coordinates transformations to write it in double null form (\ref{metricn1}). And thus one can use all the results in the last section. Then one makes inverse coordinates transformation to get the results in the original coordinates. For the special status of static spacetime, we also explicitly present the Misner-Sharp mass in the static coordinates.

  To obtain the Misner-Sharp mass, we only need $T_{tt},~T_{rr},$ and $~T_{rt}$. The useful components of the field equation in this coordinate system read,
\begin{eqnarray}
  8\pi G  T_{tt}&=& -\frac{\lambda(h-kh^2-c_2c_0^2m^2h^2-c_1c_0m^2rh^2+r^2\Lambda h^2-rh')}{r^2h^2},\nonumber\\
    8\pi G  T_{rr}&=& \frac{\lambda-kh\lambda-c_2c_0^2m^2h\lambda-c_1c_0m^2rh\lambda+r^2\Lambda h\lambda+r\lambda'}{r^2\lambda},\nonumber\\
8\pi G T_{tr}&=&0,
\end{eqnarray}
where a prime denotes the derivative with respect to $r$.

In this case, the definition of Misner-Sharp mass (\ref{2eq5}) becomes
   \be
 F\equiv dM_{eff}=A(t,r)dt+B(t,r)dr,
 \label{F1}
 \en
 where
 \bea
A(t,r) &=&\frac{V_k}{8\pi G h} r^2 (r,_{r}T_{tr}-r,_{t}T_{rr})=0, \\
B(t,r)
&=&\frac{V_k }{8\pi G \lambda} r^2(r,_{r}T_{tt}-r,_{t}T_{tr})=\frac{V_k}{8\pi Gh^2}[kh^2-h+(c_2c_0^2m^2+rc_1c_0m^2-r^2\Lambda)h^2+rh'].
\label{ABs}
\ena
 One immediately sees that $F$ in (\ref{F1}) is a closed form, since
\begin{eqnarray}
A_{,r}=B_{,t}=0.
\end{eqnarray}
 Hence, the Misner-Sharp mass is naturally well-defined for a static spacetime in the four-dimensional massive gravity as
\begin{eqnarray}
E_{eff} &=&\int B(t,r)dr + \int [A(t,r) - \frac{\partial}{\partial t}\int B(t,r)dr]dt \nonumber\\
&=& \frac{V_kr}{8\pi Gh}[(kh-1) - \frac{\Lambda r^2h}{3} + \frac{h(c_1c_0m^2r+2c_2c_0^2m^2)}{2}]\nonumber\\
&=&\frac{V_kr}{8\pi G}[\left(k-h^{ab}\partial_a r\partial_b r\right)- \frac{\Lambda r^2}{3} + \frac{(c_1c_0m^2r+2c_2c_0^2m^2)}{2}].\label{MSMG1}
\end{eqnarray}

In Refs.\cite{Cai2015,Vegh:2013sk}, the authors have found a static solution in the four-dimensional massive gravity like
\begin{equation}
\lambda(r)=\frac{1}{h(r)} = k +\frac{r^2}{l^2} -\frac{m_0}{r}+\frac{q^2 }{4r^{2} } + \frac{c_1c_0m^2}{2 }r+c_2c_0^2m^2.
\end{equation}
Therefore, the Misner-Sharp mass for this static solution is
\begin{eqnarray}
E_{eff} &=&\frac{V_k}{8\pi G}(m_0-\frac{q^2 }{4r}),
\end{eqnarray}
which is independent on $c_i(i=0,1,2)$ and like the case in the Reissner-Nordstorm solution. This Misner-Sharp mass will be a constant if the electric charge $q$ vanishes (which means that the stress energy vanishes), just like the \sch~case. This result indicates that for a shell dwelling at $r_1$ to $r_2$ ($0<r_1<r_2$) in such a  spacetime, the Misner-Sharp mass is zero. There is no energy flow in the whole spacetime, which is a result of the conservation of the stress energy of the matter field.

\section{stability problem}

 Before proceeding further about the thermodynamics of the massive gravity, we make a concise discussion on the stability problem. In general relativity a standard ADM decomposition of the metric reads,
 \be
 ds^2=-(N^2-N_iN^i)dt^2+2N_idtdx^i+\gamma_{ij}dx^idx^j,
 \en
 where $N$ denotes the lapse and $N_i$ denote the shift functions, and $\gamma_{ij}$ is the spatial metric. With such a decomposition, the Hilbert-Einstein action reads,
 \be
 S=\frac{1}{16\pi G}\int d^4x (\pi^{ij}\dot{\gamma}_{ij}+N_{\mu}R^{\mu}+\sqrt{-g}\frac{6}{l^2}),
 \label{actionadm}
 \en
 in which $(\pi^{ij}, \gamma_{ij})$ are conjugate pairs with respective to this action, $N_{\mu}=(N,~N_i)$, and
 \be
 R^0=\sqrt{\gamma} \left[\textbf{R}+\frac{1}{\gamma}(\frac{\pi^2}{2}-\pi_{ij}\pi^{ij})\right],
 \label{hami}
 \en
 \be
 R^i=2\sqrt{\gamma}~\nabla_j\left(\frac{\pi^{ij}}{\sqrt{\gamma}}\right),
 \label{shami}
 \en
  where $\gamma$ is the determinant of $\gamma_{ij}$,  $\textbf{R}$ is the three dimensional Ricci scalar of $\gamma_{ij}$. Because $\gamma_{ij}$ is symmetric, there are six propagating degree of freedoms at the most. Three of them are ordinary translational degree of freedoms of a mass point in three directions. Two of them are polarization degree of freedoms, and the residue one may become a ghost. However, $N_{\mu}$ appear in the action (\ref{actionadm}) are four lagrange multipliers, which correspond to four constraints in the resulting equations of motions. Thus, there are only two polarization degree of freedoms really $free$ in general relativity.

  The situation is different in the massive case (\ref{actionmassive}), which can be rewritten as,
  \be
  S=\frac{1}{16\pi G}\int d^4x \left(\pi^{ij}\dot{\gamma}_{ij}+N_{\mu}R^{\mu}+m^2P(N_{\mu},\gamma_{ij},f)+\sqrt{-g}\frac{6}{l^2}\right),
    \en
    where
    \be
    P=-2N\sqrt{\gamma}({\rm tr} \sqrt{f/g}-1).
    \en
    Because $f$ is a singular matrix with rank 2,  $\sqrt{f/g}$ is a rank 2 matrix. The term $-1$ in the bracket is to ensure the term $P$ does not contain a cosmological constant.  Apparently, all $N_{\mu}$ are no longer lagrange multipliers, and thus the constraints (\ref{hami}) and (\ref{shami}) are turned off.  So all the six possible degrees of freedom of $\gamma_{ij}$ are liberated, including the ghost.

     If the four $N_{\mu}$ depend only on three combinations, defined as $n_i=n_i(N_\mu)$, $N$ is separated as a lagrange multiplier, and thus the Hamiltonian constraint is turned on, which suppresses the ghost excitation. For the detailed demonstrations of this point, see \cite{hassan}. In fact it is just the case of linear Fierz-Pauli massive gravity \cite{FP1}. According to this requirement we set
  \be
  N^i=c^i(n^i,\gamma_{ij})+Nd^{i}(n^i,\gamma_{ij}).
  \label{repara}
  \en
  We emphasize that $c^i$ and $d^i$ are not functions of $N$. Further, we introduce
  \be
  \beta\triangleq(c^2)^2f_{22}+(c^3)^2f_{33},~~~~~~~~~ \sqrt{\beta}L^i_j\triangleq\sqrt{(\gamma^{im}-d^id^m)f_{mj}}~.
  \en
  Then we obtain a solution (not the general solution) for $c^i$ and $d^i$,
  \bea
  c^i&=&n^i, \\
  d^i&=&L^i_mn^m.
  \ena
  One can confirm that $N$ becomes a lagrange multiplier in the action after one makes a reparametrization of the ADM decomposition (\ref{repara}) when the above conditions are satisfied. Therefore, the ghost excitation is suppressed.

\section{Thermodynamics for the massive gravity}
There are several approaches to analyze the thermodynamics for a gravity theory~\cite{Jac,eling,pad,Cai:2008ys,hu1,Cai-Kim,CCHK,aka,Gong:2007md}. In this article, we use a direct approach, which just investigates the unified first law projected on a trapping horizon~\cite{Cai:2008ys,hu1,Cai-Kim,CCHK,aka,Gong:2007md}. For simplicity, we take it in the FRW universe into account.l

The metric of FRW universe is
\begin{equation}
ds^2 = -dt^2 + a^2(t)\Bigg(\frac{d\rho^2}{1-k\rho^2}+\rho^2(d\theta^2+\text{sin}^2\phi^2)\Bigg).
\end{equation}
After setting $r(t,\rho )\equiv a(t)\rho $, one can easily obtain
\begin{equation}
ds^2 = -dt^2 + e^{2\psi(t,\rho)}d\rho^2 + r^2(t, \rho) (d\theta^2 + \text{sin}^2\theta d\phi^2), \label{FRWnew}
\end{equation}
where $e^{\psi (t,\rho
)}=\frac{a(t)}{\sqrt{1-k\rho ^{2}}}$. For the metric (\ref{FRWnew}), the useful components in the field equation (\ref{Eqn1}) can be explicitly expressed as
\begin{eqnarray}
8\pi G T_{tt}&=& \frac{c_2c_0^2m^2+3k\rho^2+c_1c_0m^2\rho a - \Lambda \rho^2 a^2 + 3\rho^2\dot{a}^2}{\rho^2 a^2}, \notag\\
8\pi G T_{\rho\rho}&=& \frac{c_2c_0^2m^2+k\rho^2- \Lambda \rho^2 a^2+\rho^2\dot{a}^2+\rho a(c_1c_0m^2+2\rho \ddot{a})}{\rho^2 (-1+k\rho^2)},\notag\\
8\pi G T_{t\rho}&=&0,
\end{eqnarray}
and the corresponding Misner-sharp mass is
\begin{equation}
E_{eff} = \frac{r[3c_1c_0m^2r - 2\Lambda r^2 + 6(1+c_2c_0^2m^2-r_{,\rho}^2e^{-2\psi}+r_{,t}^2)]}{12}.
\end{equation}
From the unified first law, we have
\begin{equation}
dE_{eff}=A\Psi_a dx^a + WdV = A(T^b_a\partial_br + W\partial_ar)dx^a,
\end{equation}
where
\begin{eqnarray}
\Psi_a&=&\bigg[h^{\rho\rho}T_{t\rho}r_{,\rho}+\frac{1}{2}(h^{tt}T_{tt}-h^{\rho\rho}T_{\rho\rho})r_{,t}\bigg] (dt)_a+\bigg[h^{tt}T_{t\rho}r_{,t}+\frac{1}{2}(h^{\rho\rho}T_{\rho\rho}-h^{tt}T_{tt})r_{,\rho}\bigg](d\rho)_a , \nonumber\\
 W &=& -\frac{1}{2}(h^{tt}T_{tt} + h^{\rho\rho}T_{\rho\rho}). \label{Psitrho}
\end{eqnarray}

Note that, the apparent horizon of FRW universe is located at $r_A = \frac{1}{\sqrt{H^2+\frac{k}{a^2}}}$ \cite{Cai:2008ys,hu1,Cai-Kim}. A recent approach shows that the apparent horizon emits Hawking radiations in FRW universe in Einstein gravity \cite{hu1}.   Here, on the apparent horizon of a FRW universe in massive gravity, the energy crossing the
apparent horizon within time interval $dt$ is~\cite{CCHK,Cai-Kim,Cai:2008ys}
\begin{equation}
\delta Q=dE_{eff}|_{r_{A}}=A\Psi_adx^a|_{r=r_A} = A(\Psi_t-\Psi_\rho H\rho)dt.
\end{equation}
where $d\rho = \frac{1}{a}dr-H\rho dt$ has been used. After inserting the (\ref{Psitrho}) into the above equation, we obtain
\begin{equation}
\delta Q=dE_{eff}|_{r_{A}}=A\Psi_adx^a|_{r=r_A} = -Hr_A^3(-\frac{k}{a^2} - \dot{H})dt.
\end{equation}
Note that, the relationship between the entropy and horizon area just depends on the fundamental gravity theory, while the temperature just depends on the metric of spacetime. Therefore, the temperature of FRW is still $T=\frac{1}{2\pi r_A}$ \cite{Cai:2008ys,hu1,Cai-Kim}, while the entropy of apparent horizon is $S=A/4=\pi r_A^2$ \cite{Cai2015}. Hence, the Clausius relation is same as the Einstein case \cite{Cai:2008ys}
\begin{equation}
TdS = -H r_A^3(\dot{H}-\frac{k}{a^2})dt.
\end{equation}
Obviously, the usual Clausius relation
$\delta Q= TdS$ does hold on the apparent horizon of FRW universe, which indicates that the massive gravity is a equilibrium state.

 \section{conclusion and discussion}
 The Misner-Sharp mass is an interesting approach in the studies of energy in gravitational field. It is very useful to discuss  a gravitational system with energy transformation in different regions and energy conversion among different forms. Misner-sharp mass plays a critical role in the unified first law. In our paper, we derive the general form of Misner-Sharp mass via the inverse unified first law method, and confirm it by the conserved charge method. We find that the existence of a well-defined Misner-Sharp mass does not yield additional requirement of the theory, which is a distinguished feature in higher derivative theories. We demonstrate that the massive gravity with singular reference metric is ghost free. Also, for the cases of a FRW universe and a static spacetime with three types of two dimensional maximally symmetric subspaces, we present the concrete forms of the Misner-Sharp mass. Especially, we explore the thermodynamics of a FRW universe in massive gravity, and show that the massive gravity can be in thermodynamic equilibrium, i.e., there is no extra entropy production term in this case.

It should be pointed out, due to the massive graviton, the stress energy of matter field is usually not conserved if we require the massive gravity theory is diffeomorphism invariant. However, it is interesting that the stress energy of matter field is conserved in our case. Moreover, the existence of the Misner-sharp mass for the massive gravity in our case also does not lead to extra constraint, which is notable in modified gravities. Note that, all our results are just investigated under the metric ansatz (\ref{metricn1}) and the reference metric (\ref{reference}), which both have the maximally symmetric subspaces, and hence the maximal symmetry may deduces this conservation and non-extra constraint. Therefore, it is an interesting issue to find out a solution where the stress energy of matter field is not conserved. After finding the solution, we can investigate the Misner-Sharp mass and thermodynamics for this solution, which will be helpful for us to understand the energy flows between gravitational field and matter field.  In addition, it has been found that the mass of the graviton can deduce many interesting results in the holographic model via the AdS/CFT correspondence. For example, the dual conductivity generally
exhibits a Drude peak which approaches a delta function in the massless gravity limit, the mass of the graviton can play the same role as the lattice in the holographic conductor model \cite{Blake:2013owa,Ling:2014laa}, and conservation of energy-momentum in the dual field theory is usually violated, i.e., momentum dissipation and relaxation \cite{Davison:2013jba,Davison:2014lua}, etal.  Therefore, it is also another very interesting issue to have further study on the holographical massive gravity.

\section{Acknowledgements}
We thank the anonymous referee for her/his valuable suggestions. This work is supported by the Fundamental Research Funds for the Central Universities under grant No. NS2015073, National Natural Science Foundation of China (NSFC) under grant Nos. 11105004, 11075106 and 11275128, and Shanghai Key Laboratory of Particle Physics and Cosmology under grant No. 11DZ2260700. In addition, this work is also supported by the Program for Professor of Special Appointment (Eastern Scholar) at Shanghai Institutions of Higher Learning, National Education Foundation of China under grant No. 200931271104, and the Open Project Program of State Key Laboratory of Theoretical Physics, Institute of Theoretical Physics, Chinese Academy of Sciences, China (No. Y5KF161CJ1).

\end{document}